# CLUSTER DYNAMICS AND CLUSTER MASSES

DAVID MERRITT and KARL GEBHARDT
*Department of Physics and Astronomy, Rutgers University*
*Piscataway, NJ 08855*

## ABSTRACT

In spite of the promise of new techniques for constraining the mass distribution in galaxy clusters, much remains to be learned from galaxy orbital velocities. This article reviews the theory of potential estimation in hot dynamical systems like galaxy clusters. An analysis is presented of the Coma cluster, based on a sample of $\sim 1500$ galaxies with probable membership, of which $\sim 450$ have measured velocities. The Coma data are shown to be consistent with a model in which the dark matter density falls roughly as $r^{-3}$ inside of $\sim 1.5$ Mpc ($H_0 = 50$), with, perhaps, a transition to a slower falloff at larger radii. We find no significant evidence for a core in either the galaxy number densities or the mass distribution.

# 1. Introduction

Since Zwicky's[1] realization that the gravitational mass of the Coma cluster greatly exceeds the mass in stars, there has been no compelling reason to assume that the spatial distribution of the dark matter bears any relation to the density profile defined by the galaxies. But until recently, techniques for determining the gravitational potential $\Phi(r)$ have failed to yield strong, model-independent constraints on the dark matter distribution. Happily, this situation is about to change. New data from X-ray observatories (as described by K. Yamashita, H. Böhringer and R. Mushotzky at this meeting) are providing the first detailed information about the dependence of the gas temperature on position in galaxy clusters, information which will eventually translate into strong constraints on the potential. And new techniques – such as mass-mapping using gravitational lenses – are beginning to provide estimates of the matter distribution that are free of the assumptions of dynamical equilibrium and spherical symmetry (G. Soucail, J. Miralda-Escude and I. Smail, this meeting).

At the same time, the more classical approach based on measurements of discrete galaxy velocities bears re-examining due to the large data sets that are now becoming available. Here we review the theory of kinematical mass estimation in "hot" dynamical systems like galaxy clusters. We argue that much of the past work in this field has suffered from a confusion between "model fitting" and "statistical estimation." While it is easy to construct a model that is consistent, in some weak sense, with the kinematical data, such models are often highly nonunique. Most authors have based their mass estimation algorithms on the velocity dispersion profile defined by the galaxies, thus guaranteeing that their solution for $\Phi(r)$ will be degenerate. Furthermore, the estimation of quantities like $\rho(r)$, the mass density profile, is strongly "ill-conditioned" in the sense understood by statisticians,[2] meaning that the use of *ad hoc* parametrized models is guaranteed to severely bias the answer *regardless* of the quality of the data. While many of the published algorithms are very sophisticated, and a few are reasonably nonparametric, we show that most of these accomplish nothing more than an evaluation of the virial theorem, and hence tell us little that is unique or compelling about the distribution of dark matter.

Even in a precisely spherical and relaxed cluster, reasonably model-independent estimates of $\Phi(r)$ and $\rho(r)$ require large data sets as well as numerical algorithms that can use the complete information contained within a discrete set of positions and velocities. Below we analyze the Coma cluster and conclude that a fairly wide range of models is consistent with the currently existing kinematical data. In models where the galaxy velocities are close to isotropic near the cluster center, the inferred dark-matter density falls roughly as $r^{-3}$ in the inner parts of the cluster, with no strong evidence for a core. At larger radii, $r \gtrsim 1.5$ Mpc ($H_0 = 50$ km s$^{-1}$ Mpc$^{-1}$), we find marginal evidence for a slower falloff of $\rho$ with $r$, implying a total mass for the Coma cluster that could be several times larger than the standard value based on a naive application of the virial theorem. Larger kinematical samples may eventually allow us to choose between these models.

Recent work (as reviewed by M. West at this meeing) has demonstrated the prevalence of substructure and departures from dynamical equilibrium in many galaxy clusters. Here we ignore such complications, even though they might seriously bias estimates of the mass distribution as inferred from kinematical data. Our aim is to demonstrate the difficulty of the potential estimation problem even in the idealized spherical case, as a starting point for more sophisticated studies.

# 2. Potential Estimation in Hot Systems

Dynamical masses of stellar or galactic systems are usually estimated from the line-of-sight velocities of some set of luminous tracers orbiting in the overall potential. Rough estimates of the mass are often based on the virial theorem, which for a spherical system is

$$\langle v^2 \rangle = \langle r \cdot \nabla \Phi \rangle, \tag{1}$$

with $\langle v^2 \rangle$ the mean square velocity of some sample orbiting in the potential $\Phi(r)$. In a spherical system, $\langle v^2 \rangle$ is equal to three times the mean square line-of-sight velocity, independent of any assumptions about the velocity anisotropy. In nonspherical systems (which certainly include galaxy clusters), the virial theorem contains a geometrical factor that depends on intrinsic shape and orientation, both of which are typically unknown. More serious, however, is the unknown radial dependence of $\Phi$. Without some information about the relative distribution of dark and luminous components, the virial theorem places only order-of-magnitude constraints on the total mass, even in the spherical case, and says virtually nothing about the central density or scale length of the matter that determines the potential.[10,21] Most of the cluster "virial masses" quoted in the literature were derived from a form of the virial theorem which assumes that the mass is attached to, or has the same overall spatial distribution as, the galaxies. If mass does not follow light, these estimates are worth very little.

One might hope to do better by constructing the velocity dispersion profile of the galaxies, since this function contains information about the variation with radius of the kinematical quantities. But the extra information helps surprisingly little. Idealizing a galaxy cluster as a nonrotating spherical system, the Jeans equation states

$$\frac{d\Phi}{dr} = -\frac{1}{\nu}\frac{d(\nu\sigma_r^2)}{dr} - \frac{2}{r}\left(\sigma_r^2 - \sigma_t^2\right). \qquad (2)$$

Eq. (2) contains the two velocity dispersions $\sigma_r(r)$ and $\sigma_t(r)$, measured along and tangential to the radius vector. The observed velocity dispersion at every projected radius is a complicated average along the line of sight of these two intrinsic components, and contains too little information to determine both functions independently. As a result, many different $\Phi(r)$'s can be made equally consistent with an observed velocity dispersion profile by varying the assumed dependence of anisotropy on radius. This problem is widely recognized, but it is rarely emphasized just how great the indeterminacy is if no *a priori* constraints are placed on the galaxy orbits or on the distribution of the mass. Given *perfect* measurements of the surface density and velocity dispersion profiles of a set of galaxies in a spherical cluster, the central mass density is uncertain by *several orders of magnitude*, even if one imposes the reasonable constraint that $\rho(r)$ be a declining function of radius.[3] The total mass is uncertain by a smaller, but still large, factor – roughly an order of magnitude in the case of measured profiles like those in the Coma cluster. The shapes of the galaxy orbits in the more extreme models are not very likely to occur in nature, but the dependence of quantities like the mass density on the assumed kinematics is so strong that even mild departures from isotropy can imply huge changes in the mass density.

However the amount of information in a large sample of line-of-sight velocities is much greater than that contained within the velocity dispersion profile alone. For example, if one knew the maximum line-of-sight velocity at every projected radius $R$, one would have a secure limit on depth of the potential at every intrinsic radius $r = R$, namely $\Phi(r = R) \leq -v_{max}^2(R)/2$, independent of any assumptions about orbital shapes. In galaxy clusters, this approach is not very useful, since it requires measurement of the poorly-defined wings of the velocity distribution and even then only imposes a lower limit on the depth of the potential.

We can do even better by making use of the complete distribution of line-of-sight velocities. In a spherical cluster, define the "projected distribution function" $N(R, V)$ such that $N(R, V)dV$ is the surface density at $R$ of galaxies with radial velocities in the range $V$ to $V + dV$. At a given $R = R_0$, the function $N(R_0, V)$ is the so-called "line profile," the distribution of line-of-sight velocities at that projected radius. The integral of $N$ over $V$ is the surface density profile of the kinematic sample; the first moment of $N$ over $V$ is proportional to the line-of-sight rotational velocity profile (typically negligible for galaxy clusters); the second moment gives the line-of-sight velocity dispersion profile; etc. Although — remarkably, given the importance of the question — no one has yet proven mathematically that $N(R, V)$ contains enough information to uniquely determine the gravitational potential even in a spherical system, there is good reason to believe that it constrains the potential very tightly, and perhaps uniquely.[3-6] For instance, the line profiles $N(R_0, V)$ have distinctly different shapes in systems dominated by eccentric or circular orbits (Fig. 1). This fact suggests that we can determine the velocity anisotropy directly from the shapes of the line profiles, and then use the Jeans equation (1), with the known anisotropy, to infer the mass distribution.

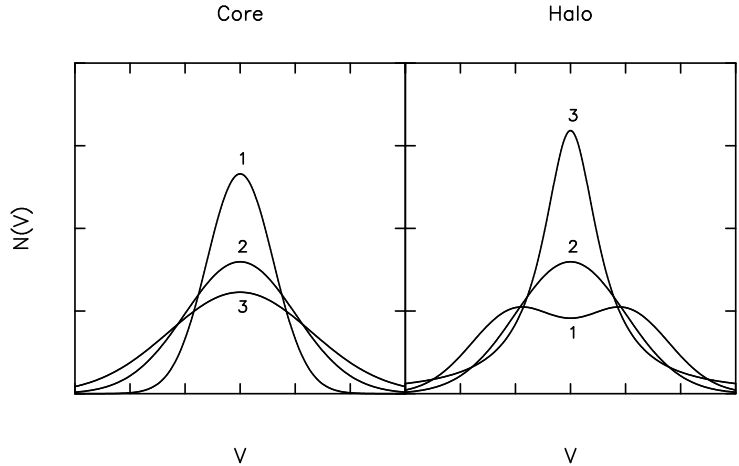

Fig. 1. Line-of-sight velocity distributions in a spherical cluster containing galaxies on three kinds of orbits. (1) Nearly circular orbits; (2) isotropic velocities; (3) strongly eccentric orbits. The left and right panels represent the appearance of the line profiles near the apparent center and in the halo, respectively.

Since the velocity dispersion profile itself tells us little about $\Phi(r)$, it follows that almost all of the kinematically-derivable information about the potential is contained within these finer details of the line-of-sight velocity distribution – crudely speaking, in the deviations of the line profiles from Gaussians (although the line profiles in even a precisely isotropic cluster need not be exactly Gaussian). This is a discouraging result, since modest departures from spherical symmetry or equilibrium in a galaxy cluster might substantially affect the shapes of these curves. In addition, rather large data sets – containing, perhaps, several hundred or even thousand velocities – are needed to construct reliable estimates of $N(R, V)$, even in a spherical cluster.

A nagging question is how to interpret past work on mass distributions in galaxy clusters as inferred from galaxy velocities. Almost all of these studies were based on the velocity dispersion profile alone – the additional (and essential) information contained within $N(R, V)$ was not used (perhaps in part because of the small size of most cluster data sets). We might expect these studies to have reached no very definite conclusions about the form of $\Phi(r)$, at least in cases where the velocity anisotropy was left as a free function to be determined by the data. On the contrary, however, many of these papers contain definite statements about the preferred form of the potential and of the galaxy kinematics – statements which appear to be justified, since the authors typically show that a particular model represents the data best. For instance, dynamical studies of the Coma cluster[7-11] often conclude that the best-fit model is one in which mass approximately follows light, and the galaxy velocities are roughly isotropic. How should we interpret these statements? Is there some objective sense in which these "best-fit" models are more likely than other models?

All of the Coma cluster studies cited above were "parametric": a set of convenient mathematical functions were postulated for representing the dynamical quantities of interest – e.g. the components of the galaxy velocity dispersion tensor,[9,10] or the phase-space density of the galaxies,[7,8,11] etc. – and the parameters of the assumed functions were then varied to maximize the goodness-of-fit of the spatially projected models to the number density and velocity dispersion profiles.

One danger of parametric techniques is that there almost always exists a single choice of parameters for which the model fits the data best, even if the underlying problem is mathematically degenerate – that is, even if the data, assumed complete and error-free, are insufficient to constrain the solution uniquely. This is because a single member of the parametric family will usually lie closest, in function space, to the region containing the set of possible, exact solutions. This single function will be selected by the optimization routine as the one that best matches the data, even though a more flexible representation of the unknown function would have yielded a range of equally-good solutions. In such cases, the "best-fit" model has no physical significance whatsoever; it is purely an artifact of the particular choice of parametric representation, since a different representation would have yielded a different "best-fit" model.

A simple experiment demonstrates the relevance of these arguments to the potential estimation problem. Suppose that we specify the observed number density and velocity dispersion profiles, $\Sigma(R)$ and $\sigma_p(R)$, of a spherical cluster. Suppose we represent the unknown distribution function describing the galaxies as a sum of basis functions, e.g.

$$f(E, L^2) = \sum_{i,j=1}^{n} c_{ij}(-E)^i L^{2j}, \qquad (3)$$

where $n$ is the number of terms retained in the expansion. By allowing $f$ to depend on the orbital angular momentum $L$ we permit the velocity distribution to be anisotropic. For any assumed potential $\Phi(r)$, we can then vary the $c_{ij}$ to optimize the fit of the projected $f$ to the measured profiles. Repeating this experiment with a family of trial potentials, we can find the pair of functions $\{\Phi, f\}$ that best reproduces $\Sigma(R)$ and $\sigma_p(R)$.

Fig. 2 shows the result for a $\Phi(r)$ defined by two parameters, core radius $R_c$ and total mass $M$. The plotted contours represent the mean square deviation of the best-fit model from the "observed" profiles, in the assumed potential. When $n$, the number of basis functions representing $f$, is small, these contours single out a particular set of values $R_c, M$ as most likely. (The lowest-order term in Eq. 3 gives the exact $f$ from which the "observed" profiles were generated; thus, the peak in Fig. 2a lies very close to the true potential.) As $n$ is increased, the $\chi^2$ contours become peculiarly elongated; when $n = 15$, there is no longer a single best-fit potential, but instead a curved region in $(R_c, M)$ space along which the goodness of fit is nearly constant. Thus, when the representation of the unknown $f$ is strongly parametrized, the algorithm gives what appears to be a unique solution for the potential; while a more flexible representation for $f$ reveals that a large number of forms for the potential are equally likely. Clearly, the "best-fit" potential in the first frame is simply an artifact; a different choice of basis set in the expansion (3) would have singled out a different point on the $(R_c, M)$-plane as optimum.

Most of the published studies of the mass distribution in galaxy clusters were based on parametrized representations with $n = 2$ or 3. As Fig. 2a shows, it is not surprising that these studies were able to find "optimum" values for the parameters defining the potential. Dejonghe[11] adopted the same series representation as in Eq. (3) for $f$, and used $n = 9$ terms in the expansion. His plots of goodness-of-fit for the Coma cluster (his figures 1 and 2) look very much like Fig. 2b here: Dejonghe also found a narrow ridge in potential parameter space along which $\chi^2$ was nearly constant, with no well-defined maximum. Ironically, Dejonghe's treatment of the Coma data, which was algorithmically much superior to the others cited above, placed the *weakest* limits on the form of the potential, since his algorithm was the most flexible and hence best able to represent the wide range of $f$'s corresponding to different potentials.

The open curve in Fig. 2c is the relation between $M$ and $R_c$ defined by the virial theorem, Eq. (1). Since the quantity $\langle v^2 \rangle$ is determined uniquely by the adopted profiles $\Sigma(R)$ and $\sigma_p(R)$, the virial theorem implies a relation between the two parameters $R_c$ and $M$ that define the adopted potential. It is apparent that the ridge of nearly-constant $\chi^2$ in Fig. 2c is simply following this virial theorem curve. In other words, almost any potential that is consistent with the virial theorem can reproduce the kinematical data equally well. Thus, one way to interpret the results of past studies of the mass distribution in galaxy clusters is to say that these authors – with the help of sometimes formidable numerical machinery – did nothing more than to evaluate the virial theorem.

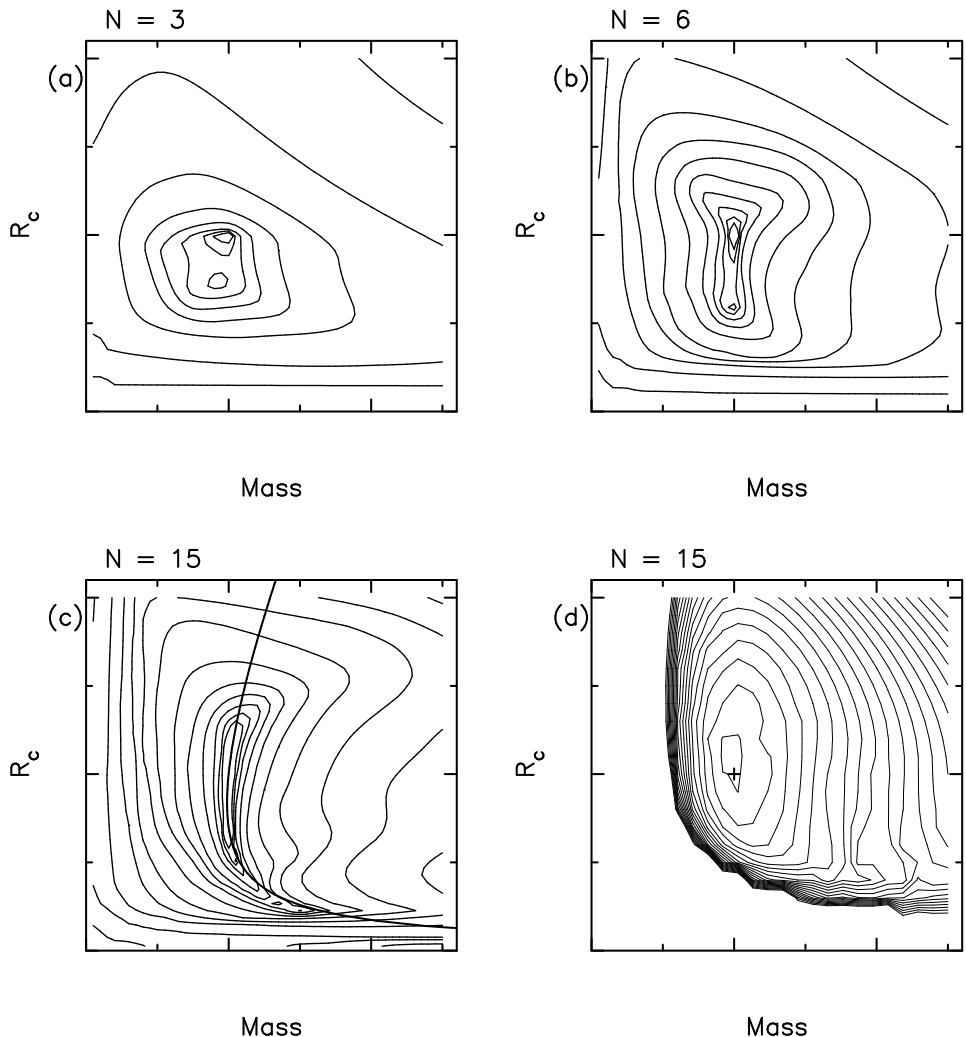

Fig. 2. Four attempts to find the "best-fit" potential from radial velocity data. In panels (a)-(c), the goodness of fit is defined as the mean square deviation of the theoretical $\Sigma(R)$ and $\sigma_p(R)$ from the data; $N$ is the number of basis functions used to approximate $f$. Panel (d) shows contours of constant likelihood, as defined in the text; the + marks the correct solution. The solid line in (c) is the curve defined by the virial theorem.

[The fact that the "best-fit" models for the Coma cluster are often characterized by a nearly constant mass-to-light ratio and by a nearly isotropic velocity distribution has a simple explanation. It so happens that the Coma data *are* well described by such a model, as first shown by Rood et al.[7] Many authors – perhaps subconciously – adopt parametrized families for functions like $f(E, L^2)$ or $\sigma_r(r)$ that contain as a special case an isotropic distribution function with a number density profile close to that of the Coma galaxies. Their optimization routine then returns this model, or one similar to it, as the best fit.]

It is possible to modify the algorithm just described to make use of the complete information contained within a discrete set of positions and velocities. One simply replaces the functional describing the goodness of fit – taken above to be the mean-square deviation of the model profiles from the observed profiles $\Sigma(R)$ and $\sigma_p(R)$ – by the *likelihood* that the particular set of $R$'s and $V$'s

would have been observed if the model were correct. That is, one varies $f$ and $\Phi$ to optimize

$$\mathcal{L} = \prod_{data} N(R_i, V_i) \tag{4}$$

where $N(R_i, V_i)$ is the value of the projected distribution function corresponding to the model $\{f, \Phi\}$ at the data point $R_i, V_i$. This modification is moderately difficult from a technical point of view, since it requires the computation of the line profiles for every considered $f$ and $\Phi$.[12] But Fig. 2d shows that the extra effort is justified: the "most likely" potential, here computed from a sample of 300 positions and velocities, is now well-defined and very close to the correct one. (The regions of zero probability in Fig. 2d correspond to potentials for which at least one of the measured velocities $V_i$ exceeds the escape velocity at $r = R_i$.) By evaluating goodness of fit via the likelihood rather than $\chi^2$, the algorithm is forced to take account of the full distribution of line-of-sight velocities, and not just the dispersions, when judging the adequacy of a model.

Although encouraging, this experiment is still parametric in its representation of $\Phi(r)$. There might easily exist some very different $\Phi(r)$, not contained within the family of functions considered, that is equally consistent with the data. In fact numerical experiments show that – while data sets of a few hundred velocities can place usefully tight constraints on a potential characterized by only two free parameters, as in the example presented above – such data can not be used to make very model-independent statements about $\Phi(r)$.[12] The reason is that many more than a few hundred velocities are needed to accurately determine the line-of-sight velocity distributions $N(R, V)$. Furthermore, features in the line profiles that one might be tempted to attribute to anisotropy may be due in a real cluster to departures from equilibrium or spherical symmetry. Perhaps the most we can hope to accomplish in galaxy clusters is to falsify some interesting, simple model – e.g. a spherical model in which mass follows light, or in which the velocity distribution is everywhere isotropic, etc. – by comparing the detailed distribution of velocities in the model with that in the observed cluster.

## 3. The Coma Cluster

A. Biviano (this meeting) describes an ongoing project to measure a large number of galaxy velocities in the Coma cluster. One motivation for this study is to look for evidence in the kinematics for substructure or departures from equilibrium. Here we present an analysis of the existing velocity data under the assumption that Coma is spherical and relaxed. We discuss the consistency of this assumption at the end.

One simple way to compute $\Phi(r)$ from the kinematical data, without assuming *ad hoc* forms for the unknown functions, is to suppose that the distribution of galaxy velocities is everywhere isotropic. The intrinsic velocity dispersion is then given by the deprojection of the observed velocity dispersion profile:

$$\nu(r)\sigma^2(r) = -\frac{1}{\pi} \int_r^\infty \frac{d(\Sigma \sigma_p^2)}{dR} \frac{dR}{\sqrt{R^2 - r^2}}, \tag{5}$$

with $\nu(r)$ the spatial density of the galaxies, obtained by deprojecting the surface density:

$$\nu(r) = -\frac{1}{\pi} \int_r^\infty \frac{d\Sigma}{dR} \frac{dR}{\sqrt{R^2 - r^2}}. \tag{6}$$

The mass within $r$ then follows from Eq. (2):

$$GM(r) = -r\sigma^2 \left( \frac{d\ln\nu}{d\ln r} + \frac{d\ln\sigma^2}{d\ln r} \right), \tag{7}$$

and the mass density is

$$\rho(r) = \frac{1}{4\pi r^2} \frac{dM}{dr}. \tag{8}$$

Here we are treating the galaxies as if they were ions in an X-ray emitting gas; the quantity $\sigma^2(r)$ plays the role of $kT(r)/m$ in the gas.

Eqs. (5) - (8) define an "inverse problem" with a unique solution $\rho(r)$, given smooth estimates of $\Sigma(R)$ and $\sigma_p(R)$. Astronomers tend to solve such problems by postulating a model, then projecting it into observable space and comparing with the data. But statisticians are fond of noting that the use of parametrized models for the solution of inverse problems is extremely dangerous, even if (unlike the case discussed in the previous section) the inverse problem has a *mathematically* unique solution. The reason is that the quantities of interest are usually related to the data via differentiations. For instance, $\rho(r)$ depends on a second derivative of $\nu(r)$ (Eqs. 7 and 8), and $\nu$ is itself a deprojection, i.e. derivative, of $\Sigma$ (Eq. 6). Any small error in the choice of parametrized model to represent $\Sigma(R)$ or $\sigma_p(R)$ will be amplified enormously in the computation of $\rho(r)$. At one level, this means simply that accurate estimation of quantities like the mass density requires high-quality data. But regardless of the quality of the data, one has a much better chance of finding the correct solution if the modeling is carried out nonparametrically. This is because a nonparametric function estimate will tend to follow the curvature implied by the data, rather than imposing a shape that is likely to be subtly wrong, even if it fits the data well in a $\chi^2$ sense.

A uniformly consistent way of solving ill-conditioned inverse problems like this one[13] is to construct smooth estimates of the input functions (here $\Sigma$ and $\sigma_p$) directly from the data, using a nonparametric algorithm, then to operate mathematically on these smooth functions to produce estimates of the functions of interest (i.e. $\Phi$, $\rho$). Confidence bands on the estimates can be constructed via the bootstrap.[14] The instability due to the inversion is dealt with by simply using a larger smoothing length on the data when constructing the estimates of $\Phi$ or $\rho$, than would be appropriate for the estimates of the data functions $\Sigma$ or $\sigma_p$ themselves.[15] By contrast, parametric techniques deal with the instability by brute force, and are almost certain to bias the answer regardless of the quality or quantity of the data.

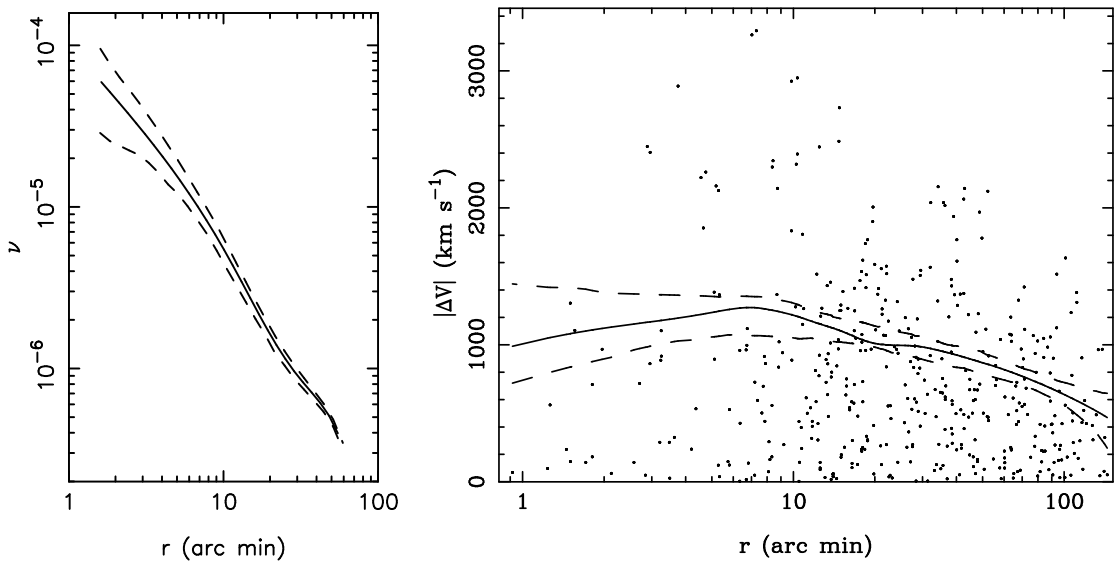

Fig. 3. Nonparametric estimates of the number density profile $\hat\nu(r)$ and line-of-sight velocity dispersion profile $\hat\sigma_p(R)$ defined by the Coma galaxies. The points in the right-hand panel are the 433 measured velocities from the compilation of T. Bird & M. King. Dashed lines are 90% confidence bands on the estimates.

Accordingly, Fig. 3 shows nonparametric estimates of $\nu(r)$ and $\sigma_p(R)$ for the Coma galaxies. The estimate $\hat\nu(r)$ was computed from a sample of 1480 galaxies identified as likely members by Mellier *et al.*,[16] using a "maximum penalized likelihood" algorithm[17] with Abell's[18] choice of cluster center. Interestingly, $\hat\nu(r)$ does not look very similar to any of the functional forms that are usually fit to it, such as the lowered isothermal sphere. Instead there is a roughly power-law cusp inside of $10'$, with $\hat{\nu} \propto r^{-1}$. The estimate $\hat{\sigma}_p(R)$ was computed from a set of 433 galaxies with measured velocities, as compiled by T. Bird and M. King. The "LOWESS" regression algorithm of W. S. Cleveland[19] was used. Both estimates are highly uncertain at radii $R \lesssim 3'$ due to the small number of bright galaxies near the center.

Fig. 4a gives the estimated mass density as a function of radius in the isotropic model, obtained by applying eqs. (7) and (8) to $\hat{\nu}$ and $\hat{\sigma}_p$. $\hat{\rho}(r)$ falls roughly as $r^{-3}$ over most of the cluster in this model, although there is a hint of a core inside of a few hundred kpc; however the 90% confidence bands are consistent with a pure power law even at small radii.

Although consistent – by construction – with the data as presented in Fig. 3, this isotropic model might still be inconsistent with the full set of velocities in Coma. Fig. 5 shows estimates of the line profiles $N(R,V)$ at three radii in the Coma cluster, as computed with an adaptive kernel algorithm[20] using galaxies from the Bird-King sample grouped in three radial annuli. Shown for comparison are the line profiles predicted by the isotropic model just discussed; the latter were computed from

$$N(R,V) = \pi \int_{R^2}^{r_{max}^2(V)} \frac{dr^2}{\sqrt{r^2-R^2}} \int_0^{-2\Phi(r)-V^2} f\left[v'^2/2 + V^2/2 + \Phi(r)\right] dv'^2, \qquad (9)$$

with the isotropic distribution function $f(E)$ computed from $\hat{\nu}$ and $\hat{\Phi}$ via Eddington's equation. (For clarity, we have omitted confidence bands on the model $N(R,V)$'s.) While the overall agreement is reasonable, there are some apparently signficant differences. The velocity distribution near the center of Coma is more peaked near $V = 0$ than in the model, perhaps indicative of an anisotropic subpopulation at the cluster center. At intermediate radii, the Coma velocity distribution is somewhat bimodal; at large radii, the distribution is once again nearly symmetric, but the mean has shifted by about 150 km s$^{-1}$ from the mean velocity near the center. Furthermore, the model profile is somewhat less peaked than the true profile at large radii.

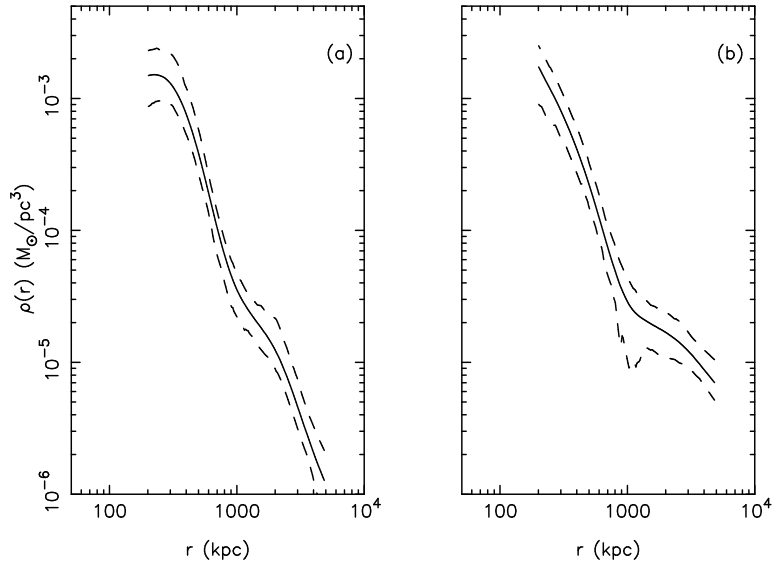

Fig. 4. Nonparametric estimates of the mass density profile $\hat{\rho}(r)$ in the Coma cluster. (a) Isotropic model; (b) anisotropic model ($r_a = 60' \approx 2.4$ Mpc). Dashed lines are 90% confidence bands on the estimates; $H_0 = 50$ km s$^{-1}$ Mpc$^{-1}$.

We can test the sensitivity of the predicted $N(R,V)$ to the assumption of isotropy by varying the assumed kinematics – taking care to leave fixed the detailed dependence of $\Sigma$ and $\sigma_p$ on radius. In other words, we vary both $\Phi(r)$, as well as the anisotropy $\beta(r) = 1 - \sigma_t^2/\sigma_r^2$, in such a way as to leave $\Sigma$ and $\sigma_p$ unchanged. One natural (though admittedly parametric) choice for describing the internal kinematics is $\sigma_r^2/\sigma_t^2 = 1 + r^2/r_a^2$, where the anisotropy radius $r_a$ defines where the galaxy velocities begin to become strongly radial. For any choice of $r_a$, one can then derive $\Phi(r)$ using a set of equations similar to those given above for the isotropic case.[21] There then exists a simple (though not unique) anisotropic distribution function $f(E + L^2/2r_a^2)$ that yields the observed $\hat{\nu}(r)$ in this potential.[22] $N(R,V)$ can be computed from this $f$ via a Monte-Carlo algorithm.

The results are shown in Figs. 4 and 5, for $r_a = 60' \approx 2.4$ Mpc. The predicted mass density profile is still well described as $\rho \propto r^{-3}$ near the cluster center, but becomes much flatter at large radii to compensate for the assumed anisotropy. (Values of $r_a$ less than about $40'$ yield a non-monotonic, and eventually negative, mass density profile; thus the model shown here with $r_a = 60'$ is close to the maximally anisotropic allowed by this form of $\beta$.) The predicted line-of-sight velocity distributions (Fig. 5) differ from those in the isotropic model only at large radii, where the preponderance of radial orbits predicts a more peaked profile, as in fact seen in the Coma data. However the predicted difference between the two models, even at large radii, is not much greater than the uncertainty in the estimates of the Coma line profiles themselves, suggesting that the current sample of $\sim 450$ velocities is barely large enough to distinguish between these two very different models. This result is not surprising: since both the area and the width of the curves in each frame of Fig. 5 are fixed by construction, the only way these curves can differ is in their detailed shapes, and large numbers of discrete velocities are required to detect such deviations with certainty.[23]

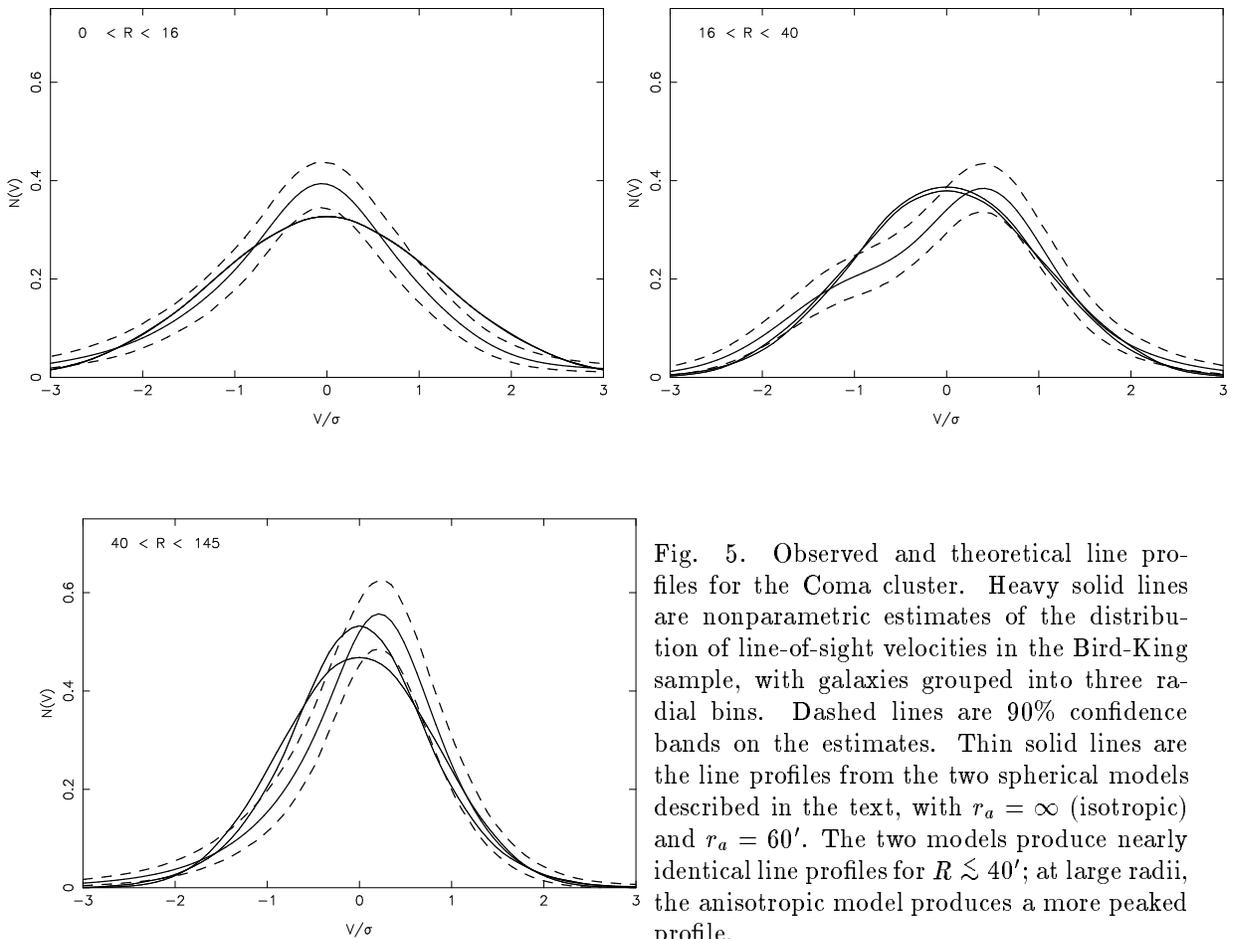

Fig. 5. Observed and theoretical line profiles for the Coma cluster. Heavy solid lines are nonparametric estimates of the distribution of line-of-sight velocities in the Bird-King sample, with galaxies grouped into three radial bins. Dashed lines are 90% confidence bands on the estimates. Thin solid lines are the line profiles from the two spherical models described in the text, with $r_a = \infty$ (isotropic) and $r_a = 60'$. The two models produce nearly identical line profiles for $R \lesssim 40'$; at large radii, the anisotropic model produces a more peaked profile.

Because $\hat{\rho}(r)$ rises more steeply into the center of Coma than the galaxy number density, the local mass-to-light ratio increases sharply at small radii in both of the models presented here, by a factor of $\sim$ 3-5 between 1 Mpc and 300 kpc.

We note that the Coma line profiles are reasonably symmetric, with no strong indications of substructure. Larger velocity samples, such as that of Biviano *et al.*, should tell us whether the slight asymmetries seen in Fig. 5 are indicative of substructure or are simply finite-sample fluctuations. (The circular orbital time exceeds $H_0^{-1}$ at $r \approx 100' \approx 4$ Mpc in our isotropic model, so we would not expect Coma to be completely relaxed at the largest radii for which we have data. But even at smaller radii, the X-ray emission[24] and galaxy positions[25] show evidence for subgroupings and ongoing formation.) The presence of dynamically-significant substructure is potentially the most serious problem facing cluster mass determinations based on galaxy velocities.

## 4. Conclusions

Even in a perfectly spherical and stationary cluster, estimation of the radial dependence of the dark matter density from line-of-sight velocity data is a hard problem. One always has the freedom to "adjust" the assumed dependence of velocity anisotropy on radius to compensate for changes in the assumed mass distribution, in such a way that the galaxy velocity dispersion profile remains precisely unchanged. But most past studies have begun by reducing all of the kinematical data to a velocity dispersion profile; the "optimum" models found in these studies can accordingly be shown to be numerical artifacts, resulting from the use of *ad hoc* parametrized functions to describe the galaxy velocity distribution function or its moments. Almost all of the information about the radial dependence of the potential is contained within the fine details of the line-of-sight velocity distribution $N(R, V)$. Furthermore, very different dynamical models – constructed so as to reproduce exactly the number density and velocity dispersion profiles defined by the galaxies – can yield very similar $N(R,V)$'s. Distinguishing between these different models, even in the idealized spherical case, requires large samples of discrete velocities, $N \gtrsim 1000$. If the cluster is nonspherical or out of equilibrium, even larger velocity samples (and more sophisticated algorithms) would be needed to constrain $\Phi(r)$ in a model-independent way. We have shown that the existing velocity data for the Coma cluster are reasonably consistent with an equilibrium model in which the mass density falls off as $\sim r^{-3}$ for $r \lesssim 1.5$ Mpc, with, perhaps, a more gradual falloff at larger radii; the total mass of the cluster is poorly defined by these data but could be several times the value derived by assuming that mass follows light.

### Acknowledgements

We are indebted to T. Bird and M. King for giving us their compilation of Coma velocities in machine-readable form.